\documentclass[doublecol]{epl2} 

\usepackage{amsmath}
\usepackage{graphicx}
\usepackage{dcolumn}
\usepackage{bm}
\usepackage{upgreek}
\usepackage{bm}
\usepackage{xcolor}

\title{Viscoelastic lubrication of a submerged cylinder sliding down an incline}
\shorttitle{Viscoelastic lubrication} 

\author{A.T. Oratis\inst{1,2} \and K. van den Berg\inst{1} \and V. Bertin\inst{1,3}  \and J.H. Snoeijer \inst{1}}
\shortauthor{A. Oratis \etal}

\institute{                    
  \inst{1} Physics of Fluids Group, Faculty of Science and Technology , University of Twente, 7500 AE Enschede, The Netherlands\\
    \inst{2} Department of Chemical Engineering, Delft University of Technology, Delft 2629 HZ, The Netherlands\\
    \inst{3} Aix Marseille University, CNRS, IUSTI UMR 7343, Marseille 13453, France
}

\abstract{
Lubrication flows between two solid surfaces can be found in a variety of biological and engineering settings.
In many of these systems, the lubricant exhibits viscoelastic properties, which modify the associated lubrication forces. 
Here, we experimentally study viscoelastic lubrication by considering the motion of a submerged cylinder sliding down an incline. 
We demonstrate that cylinders move faster when released in a viscoelastic Boger liquid compared to a Newtonian liquid with similar viscosity.
Cylinders exhibit pure sliding motion in viscoelastic liquids, in contrast to the stick-slip motion observed in Newtonian liquids.
We rationalize our results by using the second-order fluid model, which predicts a lift force on the cylinder arising from the normal-stress differences provided by the dissolved polymers.
The interplay between viscoelastic lift, viscous friction, and gravity leads to a prediction for the sliding speed, which is consistent with our experimental results for weakly viscoelastic flows.
Finally, we identify a remarkable difference between the lubrication of cylindrical and spherical contacts, as the latter does not exhibit any lift for weak viscoelasticity.}

\begin{document}

\maketitle

\section{Introduction}
Thin-film flows between solid surfaces arise in a variety of biological, environmental, and engineering settings.
These lubrication flows facilitate the relative motion of the solid boundaries by reducing their friction, as in the case of mammalian synovial joints 
\cite{mow1979mechanics} or industrial bearings and gears \cite{hamrock2004fundamentals}.
A characteristic feature of lubrication is the slenderness of the geometry, which prescribes the flow structure and induces a strong pressure within the liquid.
The build-up of pressure generates forces on the solid boundaries, such as hydrodynamic friction and normal loads. 
Lubrication flows have thus been the focus of many studies, aided by detailed analyses using long-wave expansions \cite{oron1997long}.

Considering the motion between two rigid objects with symmetric profiles, as in the case of a cylinder sliding parallel to a plane wall, the vertical lift force induced by the hydrodynamic pressure vanishes for a Newtonian viscous liquid.
The pressure adopts an antisymmetric spatial profile, leading to a zero lift force \cite{jeffrey1981slow}.
The pressure antisymmetry can be broken by introducing elasticity in the solid boundary. 
For sufficiently soft walls, the coupling between lubrication and the elastic deformation of the wall generates a lift force on the cylinder \cite{sekimoto1993mechanism,skotheim2004soft,snoeijer2013similarity,salez2015elastohydrodynamics,rallabandi2017rotation,zhang2020direct,essink2021regimes}.
The lift force leads to a larger separation with the wall and thus smaller lubrication friction.
Consequently, cylinders move faster past soft walls as compared to rigid walls \cite{saintyves2016self}.

A lift force on the cylinder can also be generated by introducing elasticity inside the lubricant, as is the case for viscoelastic liquids.
Indeed, polymers dissolved in the liquid can stretch with the flow and exert forces on the object arising from normal-stress differences \cite{morozov2015introduction}.
For instance, when a pair of closely-separated spheres sediment in a viscoelastic liquid, they get attracted to each other \cite{joseph1994aggregation,binous1999dynamic,ardekani2008two,khair2010active}.
Replacing one of the spheres with a planar wall leads to a similar effect, with the sphere being attracted to the wall during its sedimentation \cite{liu1993anomalous,becker1996sedimentation,binous1999effect,singh2000sedimentation}.
Yet, when a cylinder moves parallel to a wall in a viscoelastic fluid the attraction becomes repulsive \cite{feng1996dynamic,singh2000sedimentation}.
Why viscoelasticity causes an attraction for spheres and a repulsion for cylinders remains unclear.

In this Letter, we experimentally study the effects of viscoelastic stresses on the sliding motion of a submerged cylinder down an incline.
We show that cylinders move faster in viscoelastic liquids than in Newtonian liquids, and explore to what extent these results can be interpreted using the second-order fluid model.

\begin{figure*}[ht!]
    \centering
    \includegraphics[width=\linewidth]{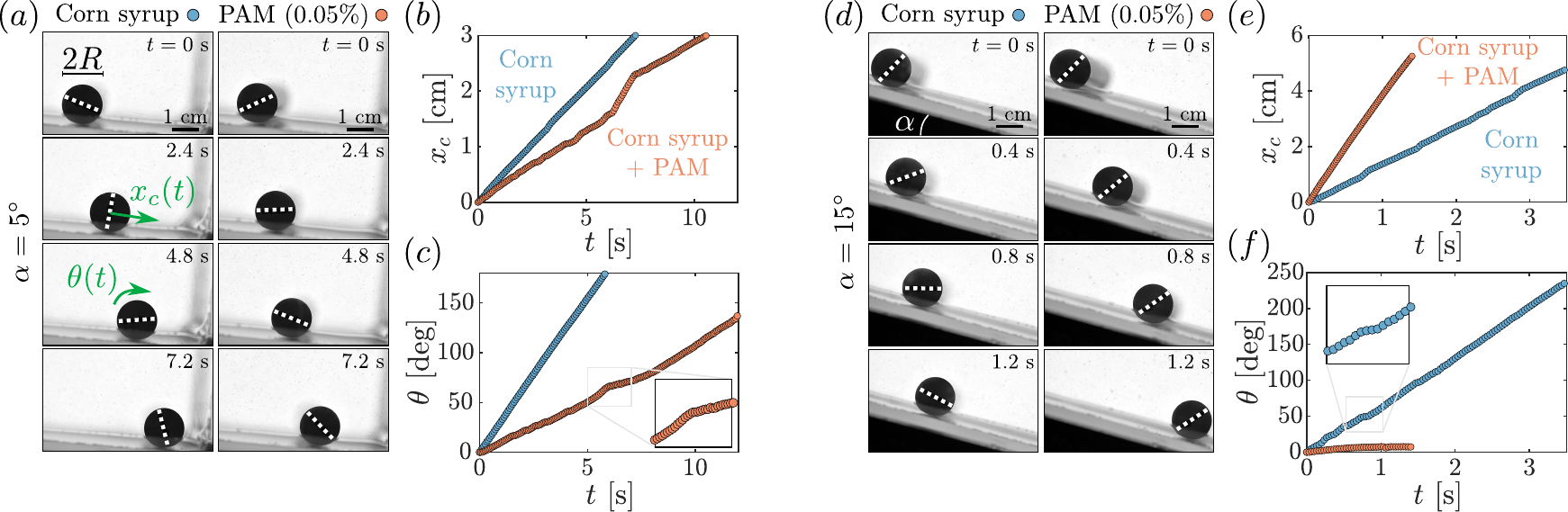}
    \caption{Sliding dynamics of a submerged steel cylinder with radius $R = 0.7$ cm.
    (\textit{a}) Experimental images showing the cylinder moving down an inclined plane with inclination angle $\alpha = 5^\circ$. 
    The cylinder rotates as it descends down the plane in both Newtonian corn-syrup solution (left) and PAM solution with $c = 0.05$ \% (right).
    (\textit{b}) The center of mass position $x_c$ against time $t$ shows that the cylinder moves faster in the Newtonian liquid (blue circles) than in the viscoelastic liquid  (red circles). The latter exhibits exhibits stick-slip motion.
	(\textit{c}) The cylinder rotation angle $\theta$ increases in time for both liquids.
	    Inset: Magnified plot of the rotation in the PAM solution  to highlight the stick-slip motion.
	(\textit{d}) Experimental images for the cylinder sliding at an inclination angle $\alpha = 15^\circ$. 
	The rotation is still apparent in the Newtonian liquid (left), but is completely suppressed in the viscoelastic liquid (right).	
    (\textit{e}) The evolution of $x_c$ indicates a much faster descent when the cylinder is immersed in the viscoelastic liquid.
    (\textit{f}) The rotation angle $\theta$ increases steadily with time for the Newtonian liquid, whereas the increase in the viscoelastic liquid is negligible.
     Inset: Magnified plot of the rotation in the Newtonian solution to highlight the stick-slip motion.}
    \label{fig:Fig1}
\end{figure*}

\section{Experimental protocol}
To study the effects of viscoelasticity on the sliding dynamics, we use a solution comprised of water, commercially available corn syrup (Chung Jung One), and Polyacrylamide (PAM $M_w = 5\times 10^6$ g/mol, Sigma Aldrich).
This combination of liquids is a common recipe used to prepare Boger fluids and probe the the effects of liquid elasticity \cite{james2009boger,keim2012fluid,tanner2016rheology,castillo2019drag,su2021coiling}.
PAM is first dissolved into water using a magnetic stirrer for 48 hours.
Five different solutions are prepared, whose PAM concentrations are $c = $ 0.10, 0.25, 0.50, 0.75, and 1.00 wt.\%.
The overlap concentration of PAM in water is approximately $c^* \approx 0.5$ wt.\% \cite{soetrisno2023concentration}, which means that the water-PAM solutions lie in the dilute/semi-dilute regime.
The water-PAM solutions are then dissolved in corn syrup at a mass ratio 10:90 using a roller bank for four days. 
A Newtonian liquid is prepared in the same manner, without dissolving PAM into water.
Each solution is measured to have the same liquid density of $\rho_\ell = 1,350$ $\rm kg/m^3$.

The rheology of every solution is characterized using a rheometer (MCR 502 with CP50-1, Anton Paar) in a cone-plate configuration by measuring the shear stress and first normal-stress difference (see Supp. Info).
As the imposed shear rate $\dot{\gamma}$ increases from 0.1 to 100 (1/s), the viscosity $\eta$ of each solution remains fairly constant, such that the viscoelastic liquids can be classified as Boger fluids \cite{james2009boger}.
The viscosity increases with the amount of dissolved polymer and varies moderately between $400 \leq \eta \leq 670$ mPa$\cdot$s.
The degree of viscoelasticity is characterized by the first normal-stress difference coefficient $\psi_1 = N_1/\dot{\gamma}^2$, where $N_1$ is the first normal-stress difference measured by the rheometer.
The resulting normal-stress difference coefficient for each solution is fitted to be $\psi_1 = \{0.5,\,2.9,\,7.2,\,13,\,13 \}$ $\times 10^{-2}$ Pa$\cdot\rm s^2$ (see Supp. Info).

\begin{figure*}[t!]
    \centering
    \includegraphics[width=0.8\linewidth]{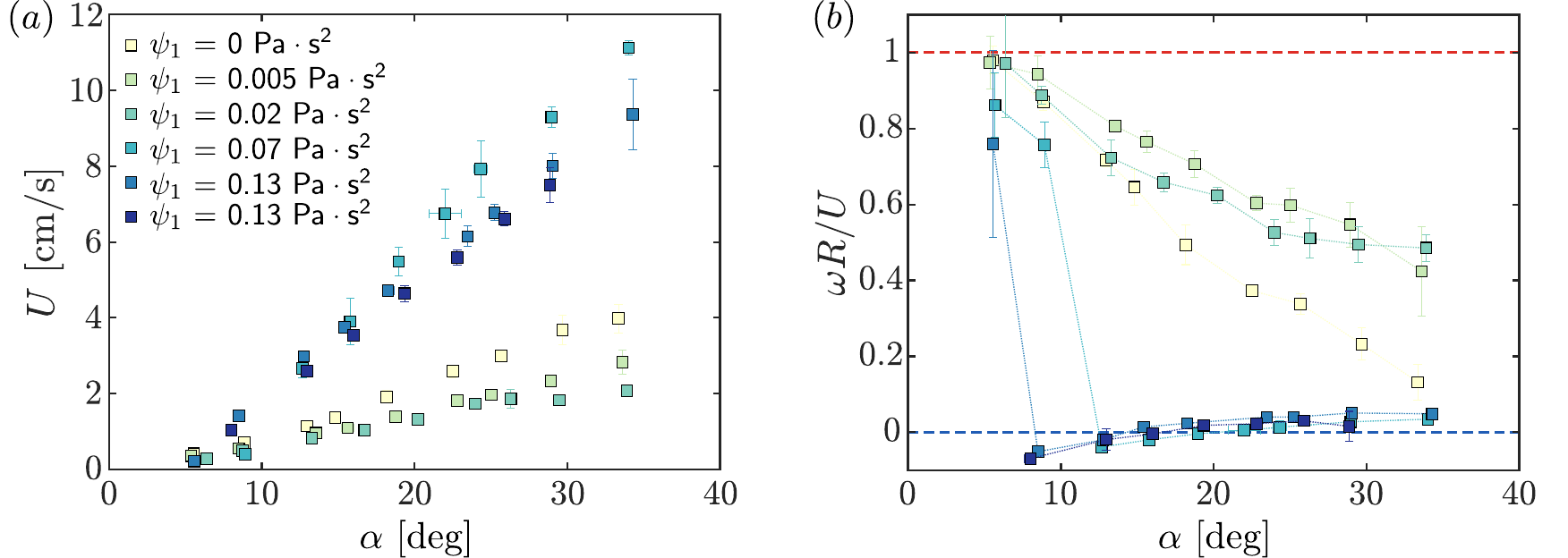}
    \caption{Effect of normal-stress differences on the steel cylinder velocity and rotation.
    (\textit{a}) The velocity of the cylinder $U$ increases with the wall inclination angle $\alpha$. 
	For large values of $\psi_1$ (blue squares), the velocity increases much more strongly with $\alpha$ compared to the Newtonian case (green squares) and low polymer concentrations (yellow squares).
    (\textit{b}) The rotational velocity $\omega R$ normalized by the cylinder velocity $U$ against the inclination angle $\alpha$. The relative rotation for the Newtonian liquid and weakly gradually decreases as the wall slope get steeper.
    Conversely, the strongly viscoelastic liquids undergo a sharp transition towards a rotation-free regime. 
    The red and blue horizontal lines depict pure rotation $\omega R/U = 1$ and pure sliding $\omega R/U = 0$ respectively.
    Each square corresponds to the average value over five runs and the error bars the standard deviation.}
    \label{fig:Fig2}
\end{figure*}

Once the liquids are prepared, they are poured in a rectangular container of dimensions 30 $\times$ 20 $\times$ 20 cm.
The container is then rotated by an inclination angle $\alpha$, which we vary in the range $5^\circ \leq \alpha \leq 35^\circ$.
A cylinder is released inside the liquid and its motion recorded with a camera (Nikon D850) at a frame rate of 60 frames per second.
We use aluminum, brass and steel cylinders, whose densities are $\rho =$ 2,590, 7,660, and 7,970 $\rm kg/m^3$ respectively.
Three different aluminum cylinders are used with radii $R = 0.6,\,1.0$, and 1.5 cm, while the radii of the brass and steel cylinders are $R = 0.4$ cm and $R = 0.7$ cm respectively.
The length of each cylinder is 10 cm, much larger than the radius, such that the system can be treated having a two-dimensional geometry. 
In addition, we also perform experiments with three steel spheres ($\rho  = 7,790$ $\rm kg/m^3$), whose radii are $R = 0.5,\,0.8$, and 1.0 cm.

\section{Cylinders}
The motion of the steel cylinder in the container for an inclination angle of $\alpha = 5^\circ$ is illustrated in Fig.~\ref{fig:Fig1}(\textit{a}).
When released inside the Newtonian liquid (i.e. corn-syrup/water solution), the cylinder exhibits a motion close to pure rolling, as inferred from the orientation of the dashed line on its cross section.
Similarly, for a viscoelastic solution with PAM concentration $c = 0.05$ wt.\%, the cylinder also rotates during its descent.
Plotting the center of mass position along the incline $x_c$ against time $t$, we observe that the cylinder moves slightly faster in the Newtonian liquid (Fig.~\ref{fig:Fig1}(\textit{b})).
The position $x_c$ appears to be increasing continuously against time.
Yet, $x_c$ increases in a much more irregular manner in the viscoelastic liquid.
The slope does not remain constant and increases at random intervals, indicative of stick-slip motion (see Supp. Video 1).
The same trends can be observed by examining the evolution of the cylinder rotation angle $\theta$ (Fig.~\ref{fig:Fig1}(\textit{c})).
The rotation angle increases steadily for the Newtonian liquid and at a faster rate than the viscoelastic solution, whose slope changes as the cylinder alternates between rolling and sliding (Fig.~\ref{fig:Fig1}(\textit{c}) inset).

Increasing the inclination angle to $\alpha = 15^\circ$ leads to a significantly different sliding motion.
While the cylinder is still rolling in the Newtonian liquid, the rotation is suppressed in the viscoelastic liquid (Fig.~\ref{fig:Fig1}(\textit{d})).
The center of mass position $x_c$ for the PAM solution no longer exhibits stick-slip and increases much faster (Fig.~\ref{fig:Fig1}(\textit{e})).
This result is surprising at first glance, as the viscoelastic solution is slightly more viscous than the Newtonian liquid.
Importantly, the dashed line on the cylinder cross section retains its orientation (Fig.~\ref{fig:Fig1}(\textit{d})).
As a result, the rotation angle $\theta$ barely increases for the viscoelastic liquid, such that the sliding is now rotation-free (Fig.~\ref{fig:Fig1}(\textit{f})).
Conversely, the rotation angle keeps increasing for the Newtonian liquid and the cylinder undergoes a stick-slip motion (see Fig.~\ref{fig:Fig1}(\textit{e}) inset and Supp. Video 2).
The higher sliding speed and absence of rotation in the viscoelastic liquid suggests a fundamental frictional change at the base of the cylinder. 

We hypothesize that a thin liquid film is entrained beneath the cylinder, removing the direct contact with the solid wall.
The absence of contact would indeed drastically reduce the sliding friction and justify the absence of rotation, as can be observed in the experimental data.
This phenomenology resembles the classical frictional transition in lubricated contacts.
Specifically, the friction coefficient can be expressed as a function of the dimensionless parameter $\eta U/N$, where $U$ is the sliding velocity and $N$ the normal load per length \cite{persson2013sliding,veltkamp2021lubricated}.
For low values of $\eta U/N$, the two surfaces are in direct solid-solid contact (boundary lubrication) and the sliding friction coefficient is large (of order 1). 
In contrast, for larger values $\eta U/N$, a continuous film lubricates the contact (hydrodynamic lubrication) and the sliding friction coefficient is largely reduced. 
An intermediate regime (mixed lubrication) separates these two cases, where there is partial contact between the surfaces. 
In the mixed lubrication regime, the sliding friction decreases with increasing speed, which is a condition to get stick-slip motion.
In our experiments, the increase of the inclination angle both increases the speed $U$ and decreases the normal load $N$. 
The combination of these effects indeed explains why we observe a transition between a friction-dominated rolling regime at low angles, to a pure sliding regime with low friction at large angles.

To systemically assess the effects of viscoelasticity on the sliding dynamics, we determine the cylinder velocity $U$ and its rotational speed $\omega$ by fitting the slope of $x_c(t)$ and $\theta(t)$ respectively.
For the cases of stick-slip motion, an average value of the slope is used.
The variation of the sliding velocity $U$ and relative rotational speed $\omega R/U$, as a function of the inclination angle $\alpha$, is plotted in Fig.~\ref{fig:Fig2}(\textit{a},\textit{b}).
The rotational speed has been normalized, such that a value of $\omega R/U = 1$ corresponds to slip-free rolling (red dashed line in Fig.~\ref{fig:Fig2}(\textit{b})), while $\omega R/U = 0$ corresponds to pure sliding (blue dashed line in Fig.~\ref{fig:Fig2}(\textit{b})).
We first focus on the Newtonian liquid (yellow squares), for which the cylinder velocity increases as the descent slope gets steeper, reaching values up to $U \approx$ 2-3 cm/s.
At low inclination angles, the relative rotation is close to unity, such that the cylinder approximately exhibits a pure rolling motion without any slip (Fig.~\ref{fig:Fig2}(\textit{b})).
As the inclination increases, the relative rotation gradually decreases, moving from slip-free rolling to stick-slip dynamics.
The lack of a pure sliding motion with increasing $\alpha$, indicates that the dynamics never reaches the lubrication regime.
There must always be some contact with the wall, suggesting a mixed lubrication, where both solid friction and hydrodynamic lubrication contribute to the sliding friction.

We now turn to the case where a small amount of polymers is dissolved in the liquid.
For the viscoelastic liquids with $\psi_1 = 0.005\,{\rm and}\,\,0.029$ Pa$\cdot\rm s^2$ (green squares), we again find an increase in the cylinder velocity with the inclination angle  (Fig.~\ref{fig:Fig2}(\textit{a})).
The cylinder velocities are slightly smaller than the Newtonian liquid, with no significant trend observed as $\psi_1$ increases.
The relative rotation in both liquids is close to unity for small inclination angles and decreases as the inclination angle becomes steeper. (Fig.~\ref{fig:Fig2}(\textit{b})).
Yet, the relative rotation decreases at a slightly lower rate than the Newtonian liquid and never reaches the pure sliding regime for the range of inclination angles tested.
We thus observe that for weakly viscoelastic liquids, the cylinder dynamics do not strongly deviate from the those observed in the Newtonian liquid.
The cylinder undergoes pure rotation or stick-slip motion with a slightly smaller velocity and higher relative rotation.
There is still contact with the wall, which means that the friction remains within the mixed lubrication regime.

Increasing the amount of dissolved polymers leads to significant changes for both $U$ and $\omega$.
For the viscoelastic liquids with  $\psi_1 = $  $0.07$, $0.13$, and $0.13$ Pa$\cdot\rm s^2$ (blue squares), the cylinder velocities remain similar to the Newtonian liquid for $\alpha < 10^\circ$.
Yet, beyond this inclination angle, the velocity values are much larger (Fig.~\ref{fig:Fig2}(\textit{a})).
The effects of viscoelasticity are also reflected in the relative rotation.
For low inclination angles, the relative rotation is high, leading to the stick-slip motion observed in Fig.~\ref{fig:Fig1}(\textit{a},\textit{b}).
As the inclination angle $\alpha$ increases, the relative rotation undergoes an abrupt transition to pure sliding.
Interestingly, the transition even leads to a slight backspin, for which the rotation is negative $\omega < 0$.  
The effect of $\psi_1$ on both $U$ and $\omega$ is not strong, with slightly higher velocities observed for the liquid with smaller $\psi_1$, suggesting a saturation of the elastic effects.
The abrupt change from stick-slip to pure sliding motions, implies a sudden transition from mixed to hydrodynamic lubrication, where a continuous thin liquid film prevents the contact between cylinder and wall.
Therefore, provided that normal-stress differences are sufficiently strong, they tend to completely suppress solid contact and lead to pure sliding motions.

\begin{figure*}[t]
    \centering
    \includegraphics[width=\linewidth]{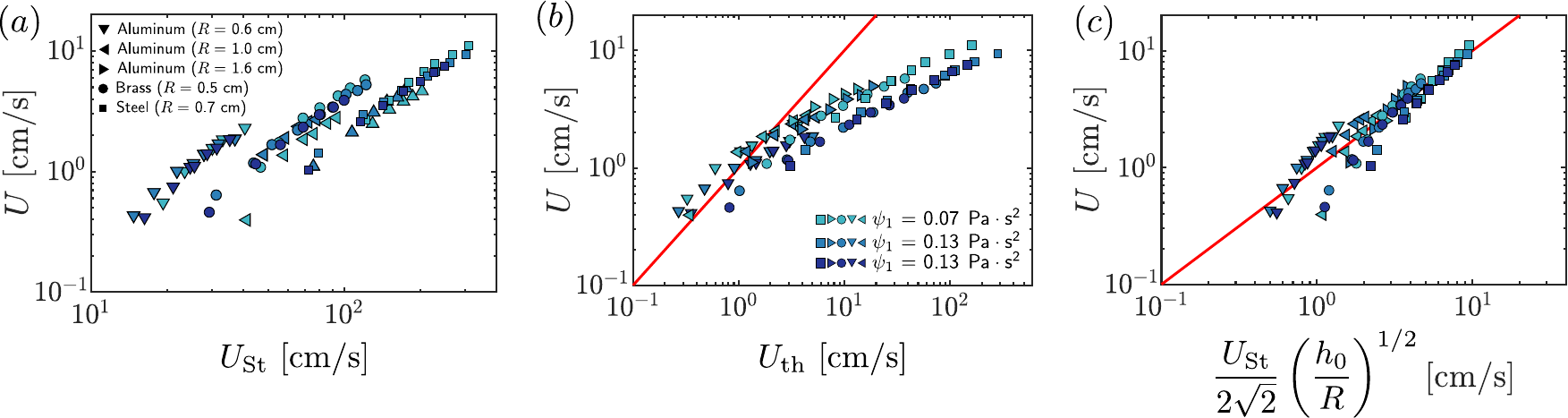}
    \caption{Different ways of rescaling the cylinder velocity $U$.
    (\textit{a}) The experimentally measured velocity $U$ against the Stokes velocity scale $U_{\rm St} = \Delta\rho g R^2\sin\alpha/\eta$. 
    The lack of collapse of the data suggests a more intricate balance that dictates the sliding velocity.
    (\textit{b}) The velocity $U$ against the $U^{\rm th}$ defined from Eq.~\eqref{eq:Uth} leads to a near collapse of the data.
    The theoretical prediction (red line) captures the smaller values of the experiments, but overestimates the velocity beyond $U \approx 2$ cm/s.
    (\textit{c}) The lubrication velocity scaling $(U_{\rm St}/2\sqrt{2})(h_0/R)^{1/2}$ leads to a better collapse of the data using a value $h_0 = 5\times 10^{-5}$ m.}
    \label{fig:Fig3}
\end{figure*}

We proceed by considering only the data of pure sliding ($\omega R/U < 0.1$) and extend our data range to the cylinders of different sizes and materials, which follow the same trend on the relative rotation as the one shown in Fig.~\ref{fig:Fig2}.
We first test how the experimental sliding speed $U$ compares to the Stokes velocity scale
\begin{equation}
U_{\rm St} = \frac{\Delta\rho g R^2\sin\alpha}{\eta}.
\label{eq:Ustokes}
\end{equation}
This velocity scale arises from a visco-gravitational balance $\Delta\rho g R^2\sin\alpha \sim \eta U$, where $\Delta\rho = \rho-\rho_\ell$ is the density difference.
Plotting the experimentally measured velocity against this prediction, we observe a scatter of the data  (Fig.~\ref{fig:Fig3}(\textit{a})).
The lack of collapse of the experimental data is not surprising, as this velocity scale does not include the effects of lubrication arising from the narrow gap separating the cylinder and the wall.
In the following we propose a mechanism that can collapse the experimentally observed velocities.

\section{Self-sustained elastic lift}
The pure sliding regime requires a mechanism of self-sustained lift preventing solid-solid contact, which can arise from the normal-stress differences in viscoelastic liquids.
To model the effects of normal-stresses, we use the second-order fluid, which is the simplest constitutive relation that admits normal-stress differences \cite{bird1987dynamics,morozov2015introduction}.
For a two-dimensional Stokes flow with velocity boundary conditions, the velocity field of a second-order fluid remains unaltered from the equivalent Newtonian  problem \cite{bird1987dynamics,tanner2000engineering}.
Hence, by solving the velocity and pressure field of the Newtonian problem, we can readily obtain the viscoelastic stresses, which are required to compute the forces exerted on the cylinder.
We consider the two-dimensional flow ($\hat{\boldsymbol{\mathrm{e}}}_x$, $\hat{\boldsymbol{\mathrm{e}}}_z$) caused by a cylinder sliding parallel to a wall at a gap distance $h_0$. 
We analyse the problem in the reference frame of the sliding cylinder, such that the wall is moving with velocity $-U$ and the flow is steady. 
The local geometry of the liquid thickness can be approximated by a parabolic profile $h(x) = h_0 [1+x^2/(2h_0 R)]$.
For a sufficiently small gap $h_0$, the lubrication approximation, $h_0/R \ll 1$, leads to a velocity field $u(x,z) = (1/2\eta)({\rm d}p_{\rm N}/{\rm d}x)(z^2 -z h) + U (z/h -1)$, where we introduced the Newtonian pressure field $p_{\rm N}(x) = 2\eta U x/h(x)^2$ \cite{jeffrey1981slow}.
Combining these two expressions with the constitutitve relation of the second order fluid, the lift force (per unit length) exerted on the cylinder can be computed analytically \cite{tanner2000engineering,oratis2024viscoelastic}, which yields the following expression
\begin{equation}
\mathcal{L} = \frac{\psi_1}{4} \int_{-\infty}^\infty \left.\left(\frac{\partial u}{\partial z} \right)^2\right|_{z=h(x)}\,\mathrm{d}x = \frac{\pi}{2\sqrt{2}}\frac{\psi_1 U^2 R^{1/2}}{h_0^{3/2}}.
\label{eq:VE_Lift}
\end{equation}
Interestingly, normal-stress differences do not generate a torque on the cylinder, nor do they modify the drag, which is given by the viscous friction $\mathcal{D} = 2\sqrt{2}\pi (\eta U R^{1/2}/h_0^{1/2})$ \cite{oratis2024viscoelastic}.

We now have all the expressions required to determine the steady sliding velocity of the cylinder.
The lift due to the normal-stress differences $\mathcal{L}$, as given by \eqref{eq:VE_Lift}, balances the weight component perpendicular to the wall 
\begin{equation}
	\frac{\pi}{2\sqrt{2}}\frac{\psi_1 U^2 R^{1/2}}{h_0^{3/2}}
	= \Delta \rho g \pi R^2 \cos\alpha.
\label{eq:Force_Balance_x}
\end{equation}
The viscous drag $\mathcal{D}$ balances the component of the weight parallel to the wall 
\begin{equation}
	2\sqrt{2}\pi \frac{\eta U R^{1/2}}{h_0^{1/2}} = \Delta \rho g \pi R^2 \sin\alpha.
\label{eq:Force_Balance_y}
\end{equation}
Solving these two equations for $U$ and $h_0$ yields 
\begin{equation}
U_{\rm th} = \frac{U_{\rm St}}{64}\left(\frac{\psi_1 \Delta \rho g R}{\eta^2}\right)\left(\frac{\sin^2\alpha}{\cos\alpha} \right),
\label{eq:Uth}
\end{equation}
\begin{equation}
h_{\rm th} = \frac{R}{512}\left(\frac{\psi_1 \Delta \rho g R}{\eta^2}\right)^2\left(\frac{\sin^2\alpha}{\cos\alpha} \right)^2.
\label{eq:hth}
\end{equation}
The theoretical sliding velocity has been expressed as the product of the Stokes velocity scale $U_{\rm St}$, defined in \eqref{eq:Ustokes}, multiplied by the geometric factor $\sin^2\alpha/\cos\alpha$ and the dimensionless viscoelastic parameter $\psi_1 \Delta \rho g R/\eta^2$.
The latter takes the form of a Weissenberg number $\psi_1 U_{\rm St} /(\eta R)$, with a strain rate given by $U_{\rm St}/R$.

The experimentally measured velocity against the theoretical prediction of \eqref{eq:Uth} is shown in Fig.~\ref{fig:Fig3}(\textit{b}).
The data collapse in a much better fashion than the Stokes scaling (Fig.~\ref{fig:Fig3}(\textit{a})).
In addition, the theoretical prediction (red line) captures the experimental data for low values of the velocity $U< 2 $ cm/s without any adjustable parameter.
For larger speeds, the prediction overestimates the cylinder velocity and the experimental data appear to have a much weaker dependence on $U_{\rm th}$.
The breakdown of our model at high velocities can possibly be attributed to two main factors.
First, the second-order fluid model provides a description only for weakly viscoelastic flows, i.e. flows at relatively low Weissenberg numbers.
For a velocity value of $U \approx 2$ cm/s, above which \eqref{eq:Uth} fails to capture the experimental data, the horizontal force balance between gravity and viscosity yields a typical thickness $h_0 \sim 10^{-5}$ m.
The corresponding Weissenberg number of the lubrication flow becomes ${\rm Wi} = \psi_1 U/(\eta h_0) \sim 100$, which is already very large.
At such values, the dependence of the normal-stresses on the shear rate might deviate from the quadratic scaling expected from the second-order fluid.
Non-linear effects, such as finite extensibility or a shear-dependent normal-stress coefficient $\psi_1(\dot{\gamma})$ are expected to come into play \cite{james2009boger}, which could lead to a saturation of the normal-stresses.
Second, the polymer solutions used are in the semi-dilute regime. 
As can be seen from the rheological measurement in the Supp. Info, the normal stress exhibits a dependence on $\dot \gamma$ that is slightly weaker than the quadratic fit used in the modeling. 

For large cylinder velocities ($U > 2$ cm/s), the experimental data do not exhibit a strong dependence on the degree of viscoelasticity, as quantified by $\psi_1$ (see Fig.~\ref{fig:Fig2}(\textit{a})).
We aim to describe the typical scaling of the cylinder speed in this regime, where the cylinder-wall contact is clearly still lubricated.
Empirically treating $h_0$ as a constant thickness, the horizontal force balance \eqref{eq:Force_Balance_x} leads to the velocity prediction $U = (U_{\rm St}/2\sqrt{2})(h_0/R)^{1/2}$.
The experimental sliding velocity agrees reasonably well with this lubrication scaling in the range $2 \leq U \leq 10 $ cm/s, showing a consistency with the lubrication theory (Fig.~\ref{fig:Fig3}(\textit{c})).
We empirically set the gap thickness to $h_0 = 5\times 10^{-5}$ m such that the best agreement is obtained.
We hypothesize that a saturation of the normal-stress differences causes the sliding velocity to transition from a regime dictated by viscoelasticity, to a regime dictated by viscous lubrication.
The exact details of the saturation of the normal-stresses and the transition to lubrication require further analysis.

\begin{figure*}[t]
    \centering
    \includegraphics[width=\linewidth]{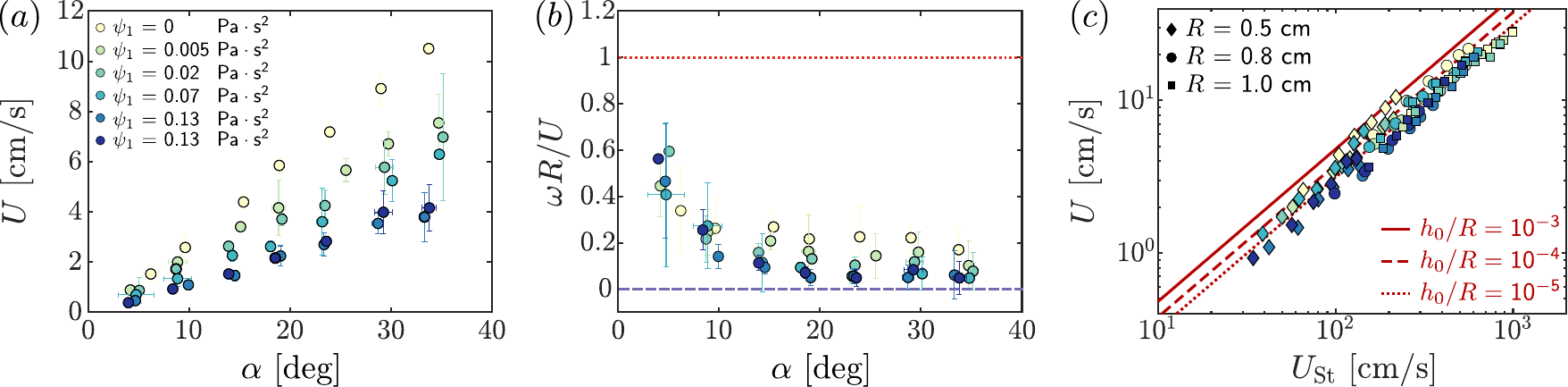}
    \caption{Sliding dynamics of a steel sphere in a viscoelastic liquid.
    (\textit{a}) The velocity $U$ for a sphere with radius $R = 0.5 $ cm against the inclination angle $\alpha$.
    As the normal-stress different coefficient $\psi_1$ increases the velocity decreases.
    (\textit{b}) The relative rotation speed $\omega R/U$ smoothly decreases as the inclination angle $\alpha$ increases. 
    The degree of viscoelasticity has a minor effect on the amount of rotation.
    (\textit{c}) The experimental velocity $U$ against the Stokes velocity $U_{\rm St}$ for different sizes (symbols) and polymer concentrations (colors).
    The theoretical prediction for Newtonian liquids is a linear line with a prefactor given by \eqref{eq:Usphere}. 
     The effect of the liquid thickness is weak as shown by the three different curves corresponding to $h_0/R = 10^{-3}\,,10^{-4}$, and $10^{-5}$.}
    \label{fig:Fig4}
\end{figure*}

\section{Spheres}
Having analyzed the dynamics of the sliding cylinder, we now extend our study to spheres.
Using the same experimental set-up, we measure the speed and rotation of a steel sphere with radius $R = 0.5$ cm.
Figure~\ref{fig:Fig4}(\textit{a}) shows the velocity $U$ against the inclination angle $\alpha$.
We observe a strong phenomenological difference with respect to the cylinders, as the addition of polymers now decreases the sliding speed, regardless of the inclination angle.
At low inclination angles, the spheres also undergo strong stick-slip motion, regardless of the amount of dissolved polymers.
However, the stick-slip motion gradually disappears with increasing inclination angles, even for the Newtonian and weakly viscoelastic liquids.
This result is also reflected in evolution of the relative rotation  $\omega R/U$ against the inclination angle $\alpha$ (Fig.~\ref{fig:Fig4}(\textit{b})).
At small inclination angles, the relative rotation is high for all liquids tested.
However, the slip-free regime ($\omega R/U = 1$) is never observed and the spheres undergo stick-slip motion (see Supp. Video 3).
Increasing the inclination angle leads to a smooth decrease of the relative rotation, which saturates to a small value near 0, for angles larger than $20^\circ$ for each liquid.
The effect of viscoelasticity on the rotation is weak, with slightly smaller rotation occurring as $\psi_1$ increases.
The absence of pure rolling motions together with the smooth transition to sliding, suggests that the sphere gradually switches from a mixed lubrication regime to a hydrodynamic lubrication regime.
These trends are markedly different from those observed for the  cylinder, where the pure sliding regime was abruptly obtained (high viscoelasticity) or never reached (low viscoelasticity).

We continue by testing how the sphere velocities compare to the Stokes velocity scale $U_{\rm St}$.
We also include the data obtained for steel spheres with radii of $R = 0.8$ cm and $R = 1$ cm.
While the Stokes velocity failed to capture the data for the cylinders (Fig.~\ref{fig:Fig3}(\textit{a})), the data for all three spheres nicely collapse on to a single curve with linear slope (Fig.~\ref{fig:Fig4}(\textit{c}))
The Stokes velocity on the horizontal axis has a higher order of magnitude than the velocities observed in our experiments, which can be attributed to the effects of lubrication.
A sphere moving parallel to a wall at a gap  distance $h_0 \ll R$, experiences a horizontal viscous drag that has been shown to follow the asymptotic expression $\mathcal{D} = 6\pi \eta U R[0.95 + (8/15) \ln(R/h_0)]$ \cite{goldman1967slow,o1967slow}.
Balancing the viscous drag with the weight component  $(4\pi/3) \Delta \rho g R^3 \sin\alpha$ yields a velocity prediction 
\begin{equation}
U = U_{\rm St}\left[ 4.29+ \frac{12}{5}\ln\left(\frac{R}{h_0} \right)\right]^{-1}.
\label{eq:Usphere}
\end{equation}
The presence of the logarithm makes the dependence of $U$ on $h_0$ weak.
To quantitatively test this prediction, we plot \eqref{eq:Usphere} using values of the dimensionless ratio $h_0/R = $ $10^{-3}$, $10^{-4}$, and $10^{-5}$, which correspond to film thickness values in the range of 0.05 to 10 microns.  
The resulting curves have small differences and are in good agreement with the experimental data (red lines in Fig.~\ref{fig:Fig4}(\textit{c})).
This agreement suggests that the decrease in velocity with $\psi_1$ is not due to the normal-stress differences but rather the increase of the solution viscosity (see Supp. Info).
Therefore, we can conclude that normal-stress differences have a limited influence on the lubrication of the spherical contact.

The absence of normal-stress effects on the sphere velocity is surprising.
Several experimental \cite{liu1993anomalous,becker1996sedimentation} and  numerical \cite{binous1999effect,singh2000sedimentation} studies have demonstrated that for large inclination angles $\alpha \approx 90^\circ$, the addition of polymers promotes the attraction of the sphere with the wall.
This attraction can be attributed to normal-stress differences, which pull the sphere towards the wall \cite{zenit2018hydrodynamic}.
A recent theoretical study by Dandekar \& Ardekani demonstrated that the hydrodynamic normal force due to normal-stress differences scales as $\mathcal{L}\sim \psi_2 U^2 R/h_0$  \cite{dandekar2021nearly}.
Here, $\psi_2$ is the second normal-stress difference coefficient, which is typically negative for polymer solutions and much smaller than $\psi_1$ \cite{bird1987dynamics}.
A negative value of $\psi_2$ would lead to a downward force and lower the sphere velocity, which rationalizes the attractive sphere-wall interaction observed numerically and experimentally for $\alpha = 90^\circ$.
Yet, even if such normal stresses were present for $\alpha < 90^\circ$, their contribution is unlikely to be significant if $\psi_2$ is small.
This interpretation is further supported by a recent experimental study, which demonstrated that normal stresses did not alter the sliding friction on curved objects \cite{veltkamp2023lubrication}.
The good agreement between the experimentally measured sliding speed and the Newtonian lubrication theory (Fig.~\ref{fig:Fig4}(c)), supports our claim that normal-stress differences have little influence in such lubricated spherical contacts.

\section{Conclusion \& outlook}
To summarize, we have investigated the dynamics of cylinders and spheres moving down an inclined plane inside a viscoelastic liquid bath.
Our experiments revealed that when submerged in highly viscoelastic liquids, cylinders adopt faster sliding speeds compared to Newtonian liquids.
The stretching of the dissolved polymers induce a lift force and push the cylinder away from the wall, which promotes the transition towards hydrodynamic lubrication and leads to smaller sliding friction as compared to Newtonian liquids.
We showed that the second-order fluid model leads to an analytical expression for the lift, which can be used to obtain a prediction for the cylinder sliding velocity.
The theoretical prediction captured well the observed sliding speed for relatively low inclination angles.
At larger angles, the prediction overestimates the sliding velocity, which is likely due to the saturation of the normal-stresses at larger Weissenberg numbers.
The polymers lead to an opposite effect for sliding spheres, whose velocity decreased with the amount of dissolved polymers.
The sliding speed was well described by Newtonian lubrication theory, suggesting that normal-stresses have a minor influence on the sphere motion.

Our experimental findings open new perspectives in viscoelastic lubrication phenomena. 
A natural extension of our work would be to simultaneously measure the film thickness and the sliding friction for viscoelastic lubricants in the hydrodynamic regime \cite{dong2023transition}.
Such experiments could elucidate how the thickness varies for the cases with large sliding velocities, and whether a transition from viscoelastic to viscous lubrication indeed exists for both cylinders and spheres.
Despite the large body of theoretical and numerical work on viscoelastic lubrication, well-controlled experiments are still needed to validate the predicted effects of viscoelasticity \cite{zenit2018hydrodynamic}.
We expect our results to provide insight into the dynamics of particle suspensions in viscoelastic liquids \cite{koch2006stress,housiadas2009rheology,rallison2012stress}, where polymer stretching complements the particle interactions to the resulting  normal stresses.

\acknowledgments
We thank C. Datt and H. Stone for stimulating discussions and R. Zenit for useful insights on the preparation of Boger fluids. This work is supported by the N.W.O through the VICI Grant No. 680-47-632.

\end{document}